\documentclass[fdp,a4paper,fleqn%
]{w-art}
\usepackage{times,cite,w-thm}
\usepackage{pst-all}
\usepackage{amsmath,amssymb}
\newcommand{\dd}{\mathrm{d}}
\newcommand{\pd}{\partial}

\newcommand{\e}{\mathrm{e}}
\newcommand{\ket}[1]{\left|#1\right\rangle}
\newcommand{\bra}[1]{\left\langle #1\right|}
\newcommand{\bracket}[2]{\left\langle%
#1\left.\right|#2\right\rangle}

\newcommand{\tr}{\mathop{\mathrm{tr}}\nolimits}

\newcommand{\V}{\mathcal{V}}

\newcommand{\Op}{\mathcal{O}}

\newcommand{\ft}[2]{{\textstyle\frac{#1}{#2}}}

\newcommand{\n}{\mathbf{n}}
\newcommand{\m}{\mathbf{m}}
\newcommand{\rr}{\mathbf{r}}
\newcommand{\s}{\mathbf{s}}

\theoremstyle{plain}

\theoremstyle{definition}

\usepackage[]{graphicx}
\begin{document}
%%    The information for the title page will be placed between
%%    \begin{document} and \maketitle. The order of most entries
%%    is determined by the class file and can not be changed by
%%    rearranging them. The maketitle command follows after the
%%    abstract.
%%
%%    Most of the following commands will be completed by the publisher.
%%
%%    The copyrightyear is defined in the .clo file as the first argument
%%    of the copyrightinfo command. If the copyrightyear differs from that
%%    value it might be adjusted by the following definition:
%%
%% \renewcommand{\copyrightyear}{2007}% uncomment to change the copyrightyear.
%%
\DOIsuffix{theDOIsuffix}
%%
%% issueinfo for the header line
\Volume{55}
\Month{01}
\Year{2007}
%%
%%    First and last pagenumber of the article. If the option
%%    'autolastpage' is set (default) the second argument may be left empty.
\pagespan{1}{}
%%
%%    Dates will be filled in by the publisher. The 'reviseddate' and
%%    'dateposted' (Published online) entry may be left empty.
\Receiveddate{XXXX}
\Reviseddate{XXXX}
\Accepteddate{XXXX}
\Dateposted{XXXX}
\keywords{Anomalous dimensions, perturbation theory, differential renormalization.}
\subjclass[pacs]{03.70.+k,11.10.Gh,11.10.Hi%%PACS-Numbers
\qquad\parbox[t][2.2\baselineskip][t]{100mm}{%
  \raggedright
  \vfill}}%

%% \pretitle{Editor's Choice}

%% We have a short and a long form for the title. The short form
%% (optional argument) goes into the running head.

\title[Dilatation Operator\& Geometry]{Dilatation Operator and Space-time Geometry}

%% Please do not enter footnotes or \inst{}-notes into the optional
%% argument of the author command. The optional argument will go into
%% the header.  If there is only one address the marker \inst{x} may be
%% omitted.

%% Information for the first author.
\author[C. Sochichiu]{Corneliu Sochichiu\inst{1,2}%
  \footnote{E-mail:~\textsf{sochichi@lnf.infn.it},
            Phone: +82\,2\,705\,7870,
            Fax: +82\,2\,704\,9031}}
\address[\inst{1}]{Center for Quantum Spacetime, Sogang University, Shinsu-dong 1, Mapo-gu, Seoul, Korea}
\address[\inst{2}]{Institutul de Fizic\u{a} Aplicat\u{a} al A\c{S}M, str. Academiei 5, Chisinau MD-2028, Moldova
}
\address[\inst{3}]{INFN-LNF, via Enrico Fermi 40,
00044 Frascati (RM), Italy.\footnote{Before April 1, 2008}}

\begin{abstract}
  We discuss the possibility of defining a dynamical model describing the RG-flow for a Quantum Field Theory. Construction of the dilatation operator is discussed in details for one-vertex one-loop level.
\end{abstract}
%% maketitle must follow the abstract.
\maketitle                   % Produces the title.

%% If there is not enough space inside the running head
%% for all authors including the title you may provide
%% the leftmark in one of the following three forms:

%% \renewcommand{\leftmark}
%% {First Author: A Short Title}

%% \renewcommand{\leftmark}
%% {First Author and Second Author: A Short Title}

%% \renewcommand{\leftmark}
%% {First Author et al.: A Short Title}

%% \tableofcontents  % Produces the table of contents.
\section{Introduction}
AdS/CFT correspondence \cite{Maldacena:1997re} conjectures an equivalence between the string/gravity theory on AdS$_5\times S^5$ and the conformal gauge theory of $\mathcal{N}=4$ super Yang--Mills (SYM) in the four-dimensional Minkowski space, which appears as the boundary to the five-dimensional anti de Sitter (AdS$_5$) space.

With the correspondence reversed, the AdS string/gravity can be regarded as built up over the four-dimen\-sional gauge theory, where the scale dependence of gauge invariant composite operators is translated into the dynamics of the dual theory \cite{Bellucci:2004fh,Sochichiu:2006uz}. From this point of view one should recover the AdS$_5\times S^5$ space as the geometry of the target space of a dynamical model whose Hamiltonian is represented by the dilatation operator. This geometrical description  is similar to one in terms of so called Connes' triples, which is used in non-commutative geometry, see e.g. \cite{Connes:1994NCG}.  According to this approach, one can describe, say, a Riemann space in terms of a triple consisting of algebra of observables, Hamiltonian and a Hilbert space. (For a brief review of this description and examples the reader is referred to \cite{Sochichiu:2002jh,Sochichiu:2005ex}.) The Hamiltonian in this approach is given by the dilatation operator, while the algebra of observables is given by the algebra of automorphisms of the algebra of composite operators\footnote{With respect to addition and multiplication.}.

Thus the Hilbert space is an important element of the description of the dynamical model. According to prescriptions of AdS/CFT correspondence, quantum states of the dual theory are represented by local gauge invariant composite operators in the gauge theory. For the linear space of such operators to be a Hilbert space one should endow it with a corresponding Hermitian metric. The main condition to be satisfied by the metric is to render the dilatation operator self-adjoint. This is required in order to have unitarity of the quantum dynamics. This condition does not fix the metric uniquely, however, it is still a non-trivial requirement. Thus the planar metric proposed in \cite{Bellucci:2004ru,Bellucci:2004qx} does not fulfill this condition for a finite $N$, but most likely it can be corrected to do so by inserting a twist operator proposed in \cite{Vafa:1994tf} (see also \cite{Verlinde:2002ig,Vaman:2002ka}).

As an approach in \cite{Bellucci:2004fh,Sochichiu:2006uz} it was proposed to interpret the letter insertion and removal operators in the SU(2) sector of $\mathcal{N}=4$ SYM as oscillator ladder operators. This interpretation leads unambiguously to a  matrix quantum mechanics with well-defined Hermitian product and a self-adjoint Hamiltonian. The Hermitian product introduced in this sector this way was equivalent to the vacuum expectation value of the product of composite operators stripped of the coordinate dependence.

Following this observation, it is tempting to generally define the Hermitian product for the dual theory basing on the vacuum expectation value of the product of composite operators. The main problem related to this approach is the ambiguity in the choice of ``canonical" letters, the orthonormal letter basis. The choice, however, may be helped by the presence of additional symmetries. Thus, in the $\mathcal{N}=4$ SYM as a conformal theory with rich symmetries one has the possibility to introduce a geometrically motivated Hermitian product based on vacuum expectation value of products of composite operators, basing on special properties of conformal theories.\footnote{For example using the fact that in conformal theories there is a direct one-to-one correspondence between local operators and quantum states.} In particular, the conformal invariance requires the pair correlators of two primary operators with different conformal weights to vanish while for those having equal one to take the form,
\begin{equation}\label{eq:op1op2}
  \langle \Op_1(x)\Op_2(0)\rangle=\frac{C_{12}}{x^{2\Delta}},
\end{equation}
where $\Delta$ is the conformal weight of the operators while $C_{12}$ is a constant. This gives the possibility to define the Hermitian product of the bra-state $\bra{\Op}$ corresponding to a composite operator $\Op^{\dag}(x)$ and a ket-state $\ket{\Op'}$ corresponding to $\Op'(x)$ in the form satisfying,
\begin{equation}
  \langle  \Op^{\dag}(x)\Op'(0)\rangle=\bra{\Op}\left(\mu^2{x^2}\right)^{-\mathbf{D}}\ket{\Op'},
\end{equation}
where $\mathbf{D}$ is the dilatation operator and $\mu$ is a mass-scale parameter. In particular, the product of states corresponding to operators $\Op_{1,2}(x)$ from \eqref{eq:op1op2} is given by $C_{12}$. Formally,
\begin{equation}\label{eq:prod-def}
  \bracket{\Op}{\Op'}=\langle \Op^{\dag}(x)\Op'(0)\rangle|_{\mu^2x^2=1},\qquad
  \bra{\Op}\mathbf{D}\ket{\Op'}=-\frac{1}{2}\mu\frac{\pd}{\pd\mu}
  \langle \Op^{\dag}(x)\Op'(0)\rangle|_{\mu^2x^2=1}.
\end{equation}

In a generic quantum field theory the correlator of two local composite operators will contain a restricted number of singular terms together with generically an infinite series of positive powers of $x$ which is the analytic part. Also in this case one may expand the correlation function in homogeneous parts in $x$. Assuming that the degree of homogeneity still corresponds to the quantum dimension of the we may hope to extend the definition \eqref{eq:prod-def} to  at least an arbitrary renormalizable theory. The checking of consistency of such definition goes beyond the scope of present note and is left for future research. Here we only assume that such a Hermitian form exists and is well-defined.

In what follows we discuss the construction of a dynamical system based on the dilatation operator for a renormalizable field theory. The discussion is based on results reported in \cite{Sochichiu:2007eu} (for a earlier discussion see also \cite{Sochichiu:2007am}).

%%%%%%%%%%%%%%%%%%%%%%%%%%%%%%
\section{The Perturbative Analysis Setup}

We start with the same setup as in \cite{Sochichiu:2007am}. According to it the local (gauge invariant) composite operators are built out of the set of (covariant) letters $\Phi_A$ as products (of traces of products) of letters. A subset of this consists of \emph{fundamental letters} which we denote $\phi_a$. They correspond to the values of the fundamental fields at the origin $\phi_a=\phi_a(0)$. The remaining letters we call \emph{derivative letters}. They are given by the values of field derivatives at the same point. In this note we are not concerned with the issue of gauge invariance, so we will use \emph{simple} derivative letters instead of covariant derivative ones,
\begin{equation}
  \Phi_A=\phi_a^{(\mathbf{n})}\equiv \pd_{(\mu_1}\dots\pd_{\mu_n)}\phi_a(0),
\end{equation}
where boldface $\mathbf{n}=\{\mu_1,\dots,\mu_n\}$ carries the multi-index of $n$ space-time indices. The parentheses denote that it is only the traceless part which is taken into account. The trace part of each letter can be removed through the equations of motion and thus would result in a redundancy of the description. One can identify the traceless multi-index label $(\mathbf{n})$ with an irreducible symmetric representation of the space-time Lorenz group.

For a theory with only massless fields of dimension one the fundamental letter Green's function is given by,
\begin{equation}\label{fund-corr}
  D_{ab}(x-y)=\rnode{C}{\phi}_a(x)\rnode{D}{\phi}_b(y)
  \ncbar[linewidth=.01,nodesep=2pt,arm=.1,angle=90]{-}{C}{D}=
  \frac{1}{4\pi^2}\frac{\delta_{ab}}{(x-y)^2}.
\end{equation}
For derivative letters we have,
\begin{equation}\label{derivative-corr}
  D^{(\n)(\m)}_{ab}(x-y)=
  \rnode{C}{\phi}^{(\n)}_a(x)\rnode{D}{\phi}^{(\m)}_b(y)
  \ncbar[linewidth=.01,nodesep=2pt,arm=.1,angle=90]{-}{C}{D}=
  \frac{(-1)^m}{4\pi^2}\pd_x^{(\n)+(\m)}\frac{\delta_{ab}}{(x-y)^2}.
\end{equation}

Generically, the product of representations $(\n)+(\m)$ contains trace parts. In this case the r.h.s. of \eqref{derivative-corr} can be presented in terms of derivatives of delta-function using the identity,
\begin{equation}
   \Box_x\frac{1}{(x-y)^2}=-4\pi^2\delta(x-y),
 \end{equation}
which is well defined as a distribution unless it is multiplied to a similar object. In contrast  the product of such functions and also of \eqref{fund-corr} is ill-defined and one needs an additional \emph{renormalization} procedure. Here we assume the \emph{differential renormalization} scheme \cite{Freedman:1991tk}.

According to the differential renormalization scheme the singular expressions like $1/(x-y)^{2k}$, $k\geq 2$ can be cast in less singular form using the following identity,
\begin{equation}\label{eq:REG}
  \left[\frac{1}{x^{2k}}\right]_{\rm reg}=-\frac{1}{4^{k-1}(k-1)!(k-2)!}
  \Box^{k-1}\frac{\ln\mu^2x^2}{x^2},
\end{equation}
where $\mu$ is a scale parameter which appears as an integration constant. It is not difficult to check by direct computation that applying all the Laplace operators to $\ln\mu^2x^2/x^2$ will formally result in $1/x^{2k}$. On the other hand the above procedure introduces a scale dependence of the regularized expression in the r.h.s. of  \eqref{eq:REG},
\begin{equation}\label{eq:SCALE}
  \mu\frac{\pd}{\pd\mu}\left[\frac{1}{x^{2k}}\right]_{\rm reg}\equiv
  -2\left[\frac{1}{x^{2k}}\right]=
  \frac{8\pi^2}{4^{k-1}(k-1)!(k-2)!}\Box^{k-2}\delta(x).
\end{equation}
These factors, which appear in the computation of anomalous dimensions, are purely ``geometric" not depending on the field theory model.

In the case of overlapping divergence the extraction of the scale factor is more evolved. For further details we refer the reader to \cite{Sochichiu:2007am}.

\section{The Evaluation of the Dilatation Operator}
According to the definition \eqref{eq:prod-def}, in order to evaluate the Dilatation Operator, we should consider the correlation function of two probe composite operators separated by $x$, which in perturbation theory can be expressed through the free correlation function,
\begin{equation}\label{eq:CORR}
  \langle \Op'(x)\Op(0)\rangle=
  \langle \Op'(x)\e^{-\int:V(\phi):}\Op(0)\rangle_0,
\end{equation}
where $\langle\dots\rangle_0$ is the free theory correlator and $V(\phi)$ is the interaction potential.\footnote{We work in Euclidean signature.}

The Wick expansion of the r.h.s. of \eqref{eq:CORR} becomes the source of singular terms which should be treated according to the differential renormalization prescription described in the previous section. Using the Wick expansion, one can express the the correlator \eqref{eq:CORR} as a free correlator of $\Op'(x)$ with a composite operator which is the result of action on $\Op(0)$ of a linear operator\footnote{Note that this is an operator acting on the linear space of composite operators.},
\begin{equation}\label{eq:LINop}
  \widehat{\e^{-\int :V:}}*\Op=\Op-\int_x :V_x:*\Op
  +\ft{1}{2!}\int_x:V_x:*\int_y :V_y:*\Op+\dots,
\end{equation}
where the (non-associative) star product is defined as,
\begin{equation}\label{eq:TWO-star}
  \Op_x(\Phi)*\Op'_y(\Phi)=
  \e^{\check{\Phi}_{Ax}D_{AB}(x-y)\check{\Phi}_{By}}
  :\Op_x(\Phi)\Op'_y(\Phi):,
\end{equation}
for two factors,
\begin{multline}\label{mul:THREE-star}
  \Op_x*\Op'_y*\Op''_z=
  \exp\{\check{\Phi}_{Ax}D_{AB}(x-y)\check{\Phi}_{By}
  +\check{\Phi}_{Ax}D_{AC}(x-z)\check{\Phi}_{Cz}\\
  +\check{\Phi}_{By}D_{BC}(y-z)\check{\Phi}_{Cz}
  \}:\Op_x\Op'_y\Op''_z:
  ,
\end{multline}
for the three factors and so on. The checked letters in \eqref{eq:TWO-star} and \eqref{mul:THREE-star} are derivatives with respect to corresponding (non-checked) letters at given point,
\begin{equation}
  \check{\Phi}_{Ax}\equiv\frac{\pd}{\pd \Phi_{Ax}}.
\end{equation}

Both \eqref{eq:TWO-star} and \eqref{mul:THREE-star}  can be obtained from the functional form of Wick expansion (see \cite{Kleinert:1996}),
\begin{equation}\label{wick-kleinert}
  \Op=:\e^{\pm\frac{1}{2}\int\dd y_1\dd y_2 \frac{\delta}{\delta \phi_{a}} D_{ab}(y_1-y_2)\frac{\delta}{\delta \phi_b}}\Op:,
\end{equation}
where $\pm$ stands for either bosons or fermions.

As the singularities and respectively the scale dependence are given by the $D$ factors which appear in the linear operator defined in eq. \eqref{eq:LINop}, the scale factor dependence of the correlator \eqref{eq:CORR} is given by the logarithmic derivative with respect to $\mu$ of the linear operator defined by \eqref{eq:LINop},
\begin{equation}
  \mathbf{D}=-\frac{1}{2}\mu\frac{\pd}{\pd\mu}
  \left[\widehat{\e^{-\int :V:}}\right]_{\rm reg}\equiv
  \left[\widehat{\e^{-\int :V:}}\right],
\end{equation}
where the square braces with subscript $[\dots]_{\rm reg}$ denotes the regularized according to differential renormalization prescription and the square braces with no subscript denotes the scale dependence of the regularized expression as given by \eqref{eq:SCALE}.

Using the perturbative expansion \eqref{eq:LINop}, the dilatation operator itself can be cast into the form,
\begin{equation}\label{Delta}
  \mathbf{D}=\int\dd y [V(y)*]+
  \frac{1}{2!}\int\dd y_1\int\dd y_2 [V(y_1)* V(y_2)*]+\dots
\end{equation}

The first term of the expansion \eqref{Delta} corresponds to one-vertex contribution, the second one to two vertex one etc.

Let us consider in more details the first two terms. The one vertex contribution can be further expanded as,
\begin{multline}\label{D:V}
  \int\dd y\, [V(y)*]
  =\int\dd y\, \left[\e^{\check{\Phi}_y\cdot D_{y}\cdot\check{\Phi}}\right]V_y\\
  =\int\dd y\left(
  \frac{1}{2}(\check{\Phi}\otimes\check{\Phi})_y\cdot
  [D_{y}\otimes D_{y}]\cdot (\check{\Phi}\otimes\check{\Phi})\right.\\
  +\left.\frac{1}{3!}(\check{\Phi}^{\otimes 3})\cdot [D_{y}^{\otimes 3}]\cdot (\check{\Phi}^{\otimes 3})
  +\frac{1}{4!}(\check{\Phi}^{\otimes 4})\cdot [D_{y}^{\otimes 4}]\cdot (\check{\Phi}^{\otimes 4})+\dots
  \right)V_y,
\end{multline}
where to further shorten the notations we introduced the following notational conventions,
\begin{align}
  \check{\Phi}_y\cdot D_{y-x}\cdot\check{\Phi}_x&=\check{\Phi}_A(y) D_{A B}(y-x)\check{\Phi}_B(x),\\
  (\check{\Phi}\otimes\check{\Phi})\cdot D\otimes D\cdot(\check{\Phi}\otimes\check{\Phi})&=
  \check{\Phi}_{A_1}\check{\Phi}_{A_2}D_{A_1B_1}D_{A_2B_2}
  \check{\Phi}_{B_1}\check{\Phi}_{B_2},\\
  \Phi^{\otimes n}&=\underbrace{\Phi\otimes\Phi\dots\otimes\Phi}_{n-\text{times}}.
\end{align}

The second line in the equation \eqref{D:V} corresponds to one-vertex one-loop contribution, the terms in the third line correspond, respectively, to one-vertex two- and three-loop contributions. In the absence of derivative letters these terms would give complete description of one-vertex contribution for a renormalizable theory. In the presence of derivative letters further terms are possible.

The one-loop part can be evaluated as follows,
\begin{multline}\label{D:V1loop}
  \frac{1}{2}\int\dd y
  (\check{\Phi}\otimes\check{\Phi})_y\cdot
  [D_{y}\otimes D_{y}]\cdot (\check{\Phi}\otimes\check{\Phi})V_y=\\
  \frac{1}{2(4\pi^2)^2}\int \dd y\sum_{\{(\m),(\n)\}}
  (-1)^{n+m}\biggl(
  (\check{\phi}_y\cdot\check{\phi}^{(\n)})
  (\check{\phi}_y\cdot\check{\phi}^{(\m)})
  \left[
  \pd^{(\n)}\frac{1}{y^2}\pd^{(\m)}\frac{1}{y^2}
  \right]+\\
  2 (\check{\phi}^{\mathbf{1}}_y\cdot\check{\phi}^{(\n)})
  (\check{\phi}_y\cdot\check{\phi}^{(\m)})
  \left[
  \pd^{(\n)+\mathbf{1}}\frac{1}{y^2}\pd^{(\m)}\frac{1}{y^2}
  \right]
  \biggr)V_y\equiv\\
  \frac{1}{2(4\pi^2)^2}
  \bigl\{
  \Delta_{(\n),(\m)}(\check{\phi}_a\check{\phi}_b(V))+
  2\Delta_{(\n)+\mathbf{1},(\m)}
  (\check{\phi}^{\mathbf{1}}_a\check{\phi}_b(V))
  \bigl\}\check{\phi}^{(\n)}_a\check{\phi}^{(\m)}_b,
\end{multline}
where we introduced the scaling factors,
\begin{equation}
  \Delta_{\s,\s'}(\V)=
  (-1)^{s+s'}
  \int_x\V_x
  \left[
  \pd^{\s}\frac{1}{x^2}\pd^{\s'}\frac{1}{x^2}
  \right],
\end{equation}
for some probe function $\V_x\equiv\V(x)$. These factors are purely geometrical, they do not depend on the model under consideration, but only on the four-dimensional space-time data. They can be evaluated to take the following form \cite{Sochichiu:2007am},
\begin{multline}\label{mul:SC-factor}
  \Delta_{(\n),(\m)}(\V)=\\
  \sum_{\substack{\rr|\n \\ \rr'|\m}}
  \frac{\pi^2n!m!}{2^{m+n-2r-1}(m+n-2r+1)!(m+n-2r)!}
  g^{\rr,\rr'}\gamma^{\n\m}_{(\n+\m-2\rr)}\\
  \times\int_x \V_x
  x^{(\n+\m-2\rr)}\Box^{m+n-2r}\delta(x)\\
  =\sum_{\substack{\rr|\n \\ \rr'|\m}}
  \frac{2\pi^2n!m!}{(m+n-2r+1)!}\
  g^{\rr,\rr'}\gamma^{\n\m}_{(\n+\m-2\rr)}
  \pd^{(\n+\m-2\rr)}\V
  .
\end{multline}
Here we use the following notations. The sum with subscript $\rr|\n$ runs through all parts $\rr$ of the index set $\n$, $\rr\subset \n$ and the coefficients $g^{\rr,\rr'}$ and $\gamma^{\n\m}_{(\n+\m-2\rr)}$ are those appearing in the expansion of the product of two traceless representations of SO(4) into traceless part and traces,
\begin{equation}\label{trtr-tr}
  x^{(\n)}x^{(\m)}=\sum_{\substack{\rr|\n \\
  \rr'|\m}}
  g^{\rr,\rr'}
  \gamma^{\n \m}_{(\m+\n-2\rr)}
  x^{2r}x^{(\m+\n-2\rr)}.
\end{equation}

Let us apply the one-vertex one-loop result \eqref{D:V1loop} to the scalar sector of $\mathcal{N}=4$ super Yang--Mills model. It is not very difficult to see that at this level the contribution to the dilatation operator comes exclusively from the scalar self-interaction which is described by the potential,
\begin{equation}
  V=\frac{g_{\rm YM}^2}{4}\tr[\phi_a,\phi_b]^2.
\end{equation}
Applying to this potential the result of \eqref{D:V1loop} and using the expression \eqref{mul:SC-factor} for the scale factors, we get,
\begin{multline}\label{comp:B}
  \mathbf{D}_{\rm 1-loop}^{\rm SO(6)}=
  \frac{1}{16\pi^2}
  \tr\left(:[\phi_a,\phi_b][\check{\phi}_a,\check{\phi}_b]:+
  :[\phi_a,\check{\phi}_b][\check{\phi}_a,\phi_b]:+
  :[\phi_a,\check{\phi}_b][\phi_a,\check{\phi}_b]:\right)\\
  =\frac{1}{8\pi^2}\tr\left(:[\phi_a,\phi_b][\check{\phi}_a,\check{\phi}_b]+
  \ft 12:[\phi_a,\check{\phi}_b][\phi_a,\check{\phi}_b]:\right),
\end{multline}
where the colon ``:'' denotes the ordering in which no checked letter acts on the letters of the same group within the colons. The equation \eqref{comp:B} is in agreement with known results \cite{Beisert:2003tq}.

\section{Discussion}

In this note we reviewed the construction of dynamical system based on RG-flows of a renormalizable Quantum Field Theory model basing on the results of \cite{Sochichiu:2007am} with the emphasis to the definition of the dynamical system. In particular, a Hermitian form rendering the dilatation operator Hermitian is proposed in terms of correlation functions of the original theory. The definition given here is somehow formal since it depends on subtraction point. However, it permits the construction in the framework of perturbation theory.

The perturbative Hermitian product we implicitly use in the construction is the one provided by the free theory. At the given level of analysis it seems to work, but in general it may require an interaction-dependent dressing in order to keep the self-adjointness of the dilatation operator. This problem is left for future study.

Another problem which was not considered here but which may appear important is the effect of fixing the physical renormalized quantities on the flow of composed operators. Again at the considered level, fixing the physical values for renormalized quantities (such as couplings or masses) plays no role. At the higher levels, however, one can not neglect this.

\begin{acknowledgement}
  This work was supported by Center for Quantum Spacetime of Sogang
  University with grant number R11 - 2005 - 021.   
  The research topic was developed with the support of European Community's Human Potential Program under contract MRTN-CT-2004-005104 ``Constituents, fundamental forces and symmetries of the universe''.
\end{acknowledgement}

% Use this code if you wish to generate your bibliography with BibTeX;
% please replace first the string "demo" below with the name(s) of
% the BibTeX data base(s) you want to use.
% The resulting bibliography-output (the contents of the .bbl file)
% must be pasted into this file before submission.
%
%% \bibliographystyle{h-elsevier}
%% \bibliography{adscft}
%
% Replace the following example bibliography with your references
% before submission:

\end{document}